# Geo-neutrinos: a new probe of Earth's interior

Gianni Fiorentini[a,b,*], Marcello Lissia[c,d], Fabio Mantovani[b,f,g], Riccardo Vannucci[h]

[a] *Dipartimento di Fisica, Università di Ferrara, I-44100 Ferrara, Italy*
[b] *Istituto Nazionale di Fisica Nucleare, Sezione di Ferrara, I-44100 Ferrara, Italy*
[c] *Istituto Nazionale di Fisica Nucleare, Sezione di Cagliari, , I-09042 Monserrato, Italy*
[d] *Dipartimento di Fisica, Università di Cagliari, I-09042 Monserrato, Italy*
[f] *Dipartimento di Scienze della Terra, Università di Siena, I-53100 Siena, Italy*
[g] *Centro di GeoTecnologie CGT, I-52027 San Giovanni Valdarno, Italy*
[h] *Dipartimento Scienze della Terra, Università di Pavia, via Ferrata 1, I-27100 Pavia, Italy*

**Abstract**

In preparation to the experimental results which will be available in the future, we study geo-neutrino production for different models of mantle convection and composition. By using global mass balance for the Bulk Silicate Earth, the predicted flux contribution from distant sources in the crust and in the mantle is fixed within a total uncertainty of ±15%. We also discuss regional effects, provided by subducting slabs or plumes near the detector. In four years a five-kton detector operating at a site relatively far from nuclear power plants can achieve measurements of the geo-neutrino signal accurate to within ±5%. It will provide a crucial test of the Bulk Silicate Earth and a direct estimate of the radiogenic contribution to terrestrial heat.

*Corresponding author. Tel.: +39-0532-974245; Fax: +39-0532-974210 *E-mail address:* fiorenti@fe.infn.it

# 1. Introduction

The nature and scale of mantle convection and the thermo-chemical evolution of Earth's mantle are still far from an appropriate understanding despite the range of observations and constraints provided by different scientific disciplines in the past half century. Arguments of mass balance and radioactive decay has lead to the canonical model of separated convective regimes with little or no mass flux between them. This paradigm has been severely challenged by mineral physics experiments, seismological observations and tomographic images, although the antagonistic model of whole-mantle convection reveals also unable to reconcile all of the geochemical and geophysical aspects.

Earth scientists now share the view that a better understanding of how the mantle really works can be achieved only by a combined approach in which all of the concepts and constraints emerging from the latest developments of formerly separate and competing disciplines are pieced together in new classes of convection models. These models can be elaborated on and tested by geodynamic, seismological, mineralogical and geochemical studies and may now include additional evidence from geo-neutrino detection.

Geo-neutrinos can be regarded as a new probe of our planet, that is becoming practical thanks to very recent and fundamental advances in the development of extremely low background neutrino detectors and in understanding neutrino propagation. Geo-neutrino detection can shed light on the sources of the terrestrial heat flow, on the present composition and on the origin of the Earth, thus providing a direct test of the Bulk Silicate Earth model and a check for non conventional models of Earth's core.

By looking at antineutrinos from nuclear reactors, the Kamioka Liquid Scintillator Anti-Neutrino Detector (KamLAND) [1] has confirmed the oscillation phenomenon previously discovered by the Sudbury Neutrino Observatory (SNO) [2] with solar neutrinos and has provided crucial information on the oscillation parameters. Since we know their destiny from production to detection, neutrinos can now be used as physical probes. Furthermore, the detector is so pure and the sensitivity is so high that KamLAND will be capable of studying geo-neutrinos, the antineutrinos originating from Earth's natural radioactivity. Indeed, from a fit to the first experimental data the KamLAND collaboration reported four events associated with $^{238}$U and five with $^{232}$Th decay chains [1]. This result provides the first insight into the radiogenic component of terrestrial heat. A new window for studying

Earth's interior has been opened and one expects more precise results in the near future from KamLAND and other detectors which are presently in preparation.

Recently, a reference model of geo-neutrino fluxes has been presented in [3]. The Reference Earth Flux model (REF) is based on a detailed description of Earth's crust and mantle and takes into account available information on the abundances of Uranium, Thorium and Potassium - the most important heat and neutrino sources - inside Earth's layers. This model has to be intended as a starting point, providing first estimates of expected events at several locations on the globe. In preparation to the experimental results which will be available in the future, from KamLAND as well as from other detectors which are in preparation, it is useful to consider geo-neutrino production in greater depth, for understanding what can be learnt on the interior of the Earth from geo-neutrino observations.

The REF model was built within the Bulk Silicate Earth (BSE) framework. The amounts of Uranium, Thorium and Potassium in the crust and in the upper mantle were derived from observational data. The content of radiogenic material in the lower part of the mantle was estimated from mass balance within BSE. We remind that BSE estimates for the total amounts of Uranium, Thorium and Potassium from different authors [4, 5, 6, 7] are quite concordant within 10%, the central values being $m_{BSE} = 0.8 \cdot 10^{17}$ kg for Uranium, $3.1 \cdot 10^{17}$ kg for Thorium, and $0.9 \cdot 10^{21}$ kg for Potassium. These values can be taken - within their uncertainties - as representatives of the composition of the present crust plus mantle system.

Different models can provide different distributions between crust and mantle, however for each element the sum of the masses is fixed by the BSE constraint. This clearly provides constraints on the geo-neutrino flux which are grounded on sound geochemical arguments. Alternatively – and this is the main point of the present paper – geo-neutrino detection can provide a test of an important geochemical paradigm.

Briefly, in this paper we shall address three questions:
(i) How sensitive are the predicted geo-neutrino fluxes to uncertainties about the mechanism of mantle circulation?
(ii) Is it possible to test the Bulk Silicate Earth model with geo-neutrinos?
(iii) Can geo-neutrino detection be sensitive to peculiar mantle structures (*e.g.* plumes)?
We shall restrict the discussion to geo-neutrinos from Uranium progeny, which are more easily detectable due their higher energy. Extension to the other chains is immediate.

In this paper, after reviewing the status of the art for neutrino detection and geo-neutrino modelling, we discuss the effect of different models for mantle structure and composition and determine the range of fluxes which are consistent with the BSE constraint. The influence of local structures of the crust and mantle is also discussed, by considering the effects of subducting slabs and of emerging plumes. The detector size needed for testing the BSE model is estimated. Our findings are summarized in the concluding remarks.

**2. State of the art**

In this section we shortly review the method for detecting anti-neutrinos and discuss the main ingredients of the reference model, providing a summary of its main predictions for geo-neutrino fluxes and event yields, referring to [8] and [3] for a more detailed presentation.

*2.1. Anti-neutrino detection*

Already in 1946 Bruno Pontecorvo [9] suggested to use nuclear reactors in order to perform neutrino experiments. Indeed, in 1953-1959 Reines and Cowan [10] showed that anti-neutrinos are real particles using nuclear reactors as a source. Since then, nuclear reactors have been extensively used to study neutrino properties. The KamLAND experiment represents the culmination of a fifty year effort, all using the same method which was applied by Reines and Cowan.

The inverse $\beta$-decay reaction, $\bar{\nu}_e + p \rightarrow e^+ + n$ (where $\bar{\nu}_e$ and p in the left side are the anti-neutrino and proton, respectively, $e^+$ and n in the right side denote the neutron and positron, respectively), is used to detect $\bar{\nu}_e$'s with energies above 1.8 MeV in liquid scintillator. The prompt signal from the positron and the 2.2 MeV $\gamma$-ray from neutron capture on a proton in delayed coincidence provide a powerful tool for reducing background and to reveal the rare interaction of antineutrinos (Fig. 1). The primary goal of KamLAND was a search for the oscillation of $\bar{\nu}_e$'s emitted from distant power reactors. The long baseline, typically 180 km, enabled KamLAND to address the oscillation solution of the 'solar neutrino problem' using reactor anti-neutrinos. KamLAND has been able to measure the oscillation parameters of electron anti-neutrinos, by comparing the observed event spectrum with that predicted in the absence of oscillation. In addition, KamLAND was capable to extract the signal of geo-neutrinos from $^{238}$U and $^{232}$Th. Due to the different energy spectra, events from Uranium and Thorium progenies can be separated. The best fit attributes 4 events to $^{238}$U and 5 to $^{232}$Th. According to [1] and [11], this corresponds to about 40 TW radiogenic heat generation, values from 0 to 110 TW being allowed at 95% C.L.

*2.2. The reference model of geo-neutrino production*

The main sources of heat and antineutrinos in the Earth's interior are Uranium, Thorium and Potassium. Through its decay chain, each nuclide releases energy together with anti-neutrinos (Table 1). From the distribution of these elements in the Earth one can thus estimate both radiogenic heat flow and the anti-neutrino flow.

The argument of geo-neutrinos was introduced by Eder [12] in the 60's and it was extensively reviewed by Krauss et al. [13] in the 1980's. Raghavan et al. [11] and Rothschild et al. [14] remarked on the potential of KamLAND and Borexino for geo-neutrino observations. The relevance of geo-neutrinos for determining the radiogenic contribution to Earth's heat flow [15] has been discussed in [3, 16, 17].

Recently, a reference model of geo-neutrino fluxes has been presented in [3]. The main ingredients of this model and its predictions for geo-neutrino fluxes and event yields are reviewed in the following. Concerning the crust, the 2°x2° model of Ref. [18] was adopted. World-averaged abundances of radiogenic elements have been estimated separately for oceans, the continental crust (subdivided into upper, middle and lower sub-layers), sediments, and oceanic crust. Although this treatment looks rather detailed on the globe scale, the typical linear dimension of each tile is of order 200 km, so that any information on a smaller scale is essentially lost. The Preliminary Reference Earth Model [19] was used for the mantle density profile, dividing Earth's interior into several spherically symmetrical shells corresponding to seismic discontinuities. Concerning its composition, a two-layer stratified model was used: for present day upper mantle, considered as the source of MORB, mass abundances of 6.5 and 17.3 ppb for Uranium and Thorium respectively and 78 ppm for Potassium were assumed down to a depth $h_0$ = 670 km. These abundances were obtained by averaging the results of Refs. [20] and [21]. Abundances in the lower mantle were inferred by requiring that the BSE constraint is globally satisfied, thus obtaining 13.2 and 52 ppb for U and Th respectively, 160 ppm for K.

From the knowledge of the source distributions, one can derive the produced antineutrino fluxes[1]:

$$\Phi_X(\vec{r}) = \frac{n_X}{4\pi\mu_X\tau_X} \int_{V_\oplus} d^3r' \frac{\rho(\vec{r}')a_X(\vec{r}')}{|\vec{r}-\vec{r}'|^2} \quad (1)$$

where the suffix $X$ denotes the element, $\tau$ is its lifetime, $\mu$ is the atom mass and $a$ is the element abundance; $n$ is the number of antineutrinos per decay chain, the integral is over the Earth's volume and $\rho$ is the local density; $(\vec{r})$ and $(\vec{r}')$ indicate the detector and the source position, respectively. The produced fluxes at several sites on the globe have been calculated within the reference model, see [3]. We concentrate here on a few locations of specific interest:

(i) For the Kamioka mine, where the KamLAND detector is in operation, the predicted uranium flux is $\Phi_U = 3.7 \cdot 10^6$ cm$^{-2}$s$^{-1}$, the flux from Thorium is comparable and that from Potassium is fourfold. Within the reference model, about 3/4 of the flux is generated from material in the crust and the rest mainly from the lower mantle.

(ii) At Gran Sasso laboratory, where Borexino [22] is in preparation, the prediction is $\Phi_U = 4.2 \cdot 10^6$ cm$^{-2}$s$^{-1}$, this larger flux arising from a bigger contribution of the surrounding continental crust. Thorium and Potassium fluxes are found to be correspondingly rescaled.

(iii) At the top of Himalaya, a place chosen so that the crust contribution is maximal, one has $\Phi_U = 6.7 \cdot 10^6$ cm$^{-2}$s$^{-1}$. The crust contribution exceeds 90%.

(iv) At Hawaii, a site which minimizes the crust contribution, the prediction is $\Phi_U = 1.3 \cdot 10^6$ cm$^{-2}$s$^{-1}$, originated mainly from the mantle.

From the produced fluxes, together with the knowledge of neutrino propagation (*i.e.* the oscillation parameters), the interaction cross section and the size of the detector, one can compute the expected event yields. These are shown over the globe in Fig. 2 (see http://www.neogeo.unisi.it/fabio/index.asp for more information). In summary, this reference model has to be intended as a starting point, providing first estimates of expected fluxes and events. In view of the present debate about mantle circulation and composition, a more general treatment is needed, which encompasses both geochemically and geophysically preferred models.

**3. Geochemistry, geophysics and geo-neutrinos**

The composition and circulation inside Earth's mantle is the subject of a strong and so far unresolved debate between geochemists and geophysicists. Geochemical evidence has been used to support the existence of two compositionally distinct reservoirs in the mantle, the borders between them being usually placed at a depth near $h_0 = 670$ km, whereas geophysics presents evidence of mantle convection extending well beyond this depth. If this convection involves the whole mantle, it would have destroyed any pre-existing layering, in conflict with geochemical evidence.

More generally, new views on mantle convection models overcome the widely diffused model of two-layer mantle convection, namely an outgassed and depleted upper layer overlying a deeper, relatively primordial and undegassed mantle layer. The ensemble of geochemical and geophysical evidence along with terrestrial heat flow-heat production balance argues against both whole mantle convection and layering at 670 km depth models, suggesting the existence of a transition between the two reservoirs (outgassed and depleted – degassed and primordial) at 1600–2000 km depth [23, 24, 25]. In the numerical simulation of their mantle convection model, Kellogg et al. [24] located this boundary at ~1600 km depth and calculated for the layers depleted and enriched in heat-producing elements a U concentration of 7 and 25.6 ppb, respectively.

In this section we look at the implications of this debate on the predicted geo-neutrino fluxes. One can build a wide class of models, including the extreme geochemical and geophysical models, in terms of just one free parameter, the depth $h$ marking the borders between the two hypothetical reservoirs:

i) Estimates of U in depleted upper mantle after crust extraction confine previously proposed values in the range of 2 to 7.1 ppb [26, 27, 28]. Given the uncertainty on these values, we assumed in a previous contribution that the uppermost part of the mantle has an average value of 6.5 ppb [3]. This value, close to the more recent consensus values of 4-5 ppb [26, 29] is here assumed, for consistency, to represent Uranium abundance ($a_u$) down to an unspecified mantle depth $h$. As shown below, the assumption of lower U abundance for the uppermost depleted mantle has limited effects on geo-neutrino flux predictions.

---

[1] We remark that angle-integrated fluxes are relevant for the non-directional geo-neutrino detection.

ii) Below $h$ we determine abundances ($a_l$) by requiring mass balance for the whole Earth. This means that Uranium mass below the critical depth, $m_{>h}$, is obtained by subtracting from the total BSE estimated mass ($m_{BSE}$) the quantity observationally determined in the crust ($m_c$) and that contained in the mantle above $h$ ($m_{<h}$):

$$m_{>h} = m_{BSE} - m_c - m_{<h} \qquad (2)$$

The abundance in the lower part is then calculated as the ratio of $m_{>h}$ to Earth's mass below $h$ ($M_{>h}$):

$$a_l = m_{>h}/M_{>h} \qquad (3)$$

This class of models, described in Figs. 3 and 4, includes a fully mixed mantle, which is obtained for $h = 25$ km (*i.e.* just below a mean crust thickness obtained averaging the vales for continental and oceanic crust) so that the strongly impoverished mantle has a negligible thickness. The traditional geochemical model corresponds to $h = h_0$. As $h$ increases, the depleted region extends deeper inside the Earth and - due to mass balance - the innermost part of the mantle becomes richer and closer in composition to the primitive mantle. These simplified models imply a uniform composition of the considered mantle shells, against the ample evidence of large regional chemical and isotopic heterogeneities. A similar argument holds for the heterogeneity in the density distribution in the Earth's interior that may also affect neutrino flux [30]. However, the choice of a gross average of compositional and density parameters is a reasonable approximation for a precise determination of the geo-neutrino fluxes, if uncertainties resulting from the neglected regional fluctuations are further evaluated (see section 5).

Let us discuss in detail a few cases, remembering that the BSE estimate for Uranium in the whole Earth is $m_{BSE} = 0.8 \cdot 10^{17}$ kg and that the best estimate for the amount in the crust [3] is $m_c = 0.35 \cdot 10^{17}$ kg so that Uranium in the mantle is expected to be $m_m = 0.45 \cdot 10^{17}$ kg.

a) In the fully mixed model, this quantity has to be distributed over the mantle mass $M_m = 4.0 \cdot 10^{24}$ kg, which yields a uniform mantle abundance $a$ = 11.25 ppb. We shall refer to this model as MIX.

b) If we keep the estimated abundance in the uppermost part ($a_u$ = 6.5 ppb) down to $h_0$ one has the REF model [3].

c) Among all possible models, the case $h$ = 1630 km is particularly interesting. Below this depth the resulting Uranium abundance is 20 ppb, corresponding to the BSE estimate. The innermost part of the mantle is thus primitive in its trace element composition and the crust enrichment is obtained at expenses of the mantle content above $h$. We shall refer to this model as PRIM.

Concerning geo-neutrino fluxes from the mantle, all the models proposed above have the same amount of heat/anti-neutrino sources and only the geometrical distribution is varied. The largest flux corresponds to the model with sources closest to the surface, *i.e.* to the MIX model. On the other hand, the minimal prediction is obtained when the sources are concentrated at larger depth, which corresponds to the PRIM case. From Table 2, the difference between the extreme cases is 8%, model REF being in between.

The abundance in the upper reservoir $a_u$ can also be treated as a free parameter. If we use an extremely low value $a_u$ = 2 ppb [27] down to about 1200 km and primitive abundance below, we obtain the minimal prediction $0.86 \cdot 10^6$ cm$^{-2}$s$^{-1}$.

We conclude this section with the following remarks:

a) Uncertainties on the geometrical distribution of trace elements in the mantle can change the REF prediction for the mantle by at most ±8%.

b) A geo-neutrino detector at a site where the contribution from the mantle is dominant (*i.e.* far from the continental crust) can be sensitive to the mantle compositional geometry only if measurements can be accurate within to the percent level.

c) Since at Kamioka mine or at Gran Sasso the mantle contribution to the total flux is about one quarter of the total [3], uncertainties on the mantle geometry imply an estimated error of about 2% on the total flux predicted with REF.

In our modelling we assumed that the Earth's core does not contain a significant amount of radioactive elements. We are aware that some authors proposed that the core is hosting some radioactive elements, and particularly K, in order to offer an alterative explanation either for the energy needed to run the Geodynamo or as a way to explain Earth's volatile elements inventory [31]. However, the proposed models of the core's energy budget imply a variety of assumptions and are vastly different, thus reaching in cases opposite conclusions, whereas geochemical evidence is in favour of a general absence of radioactive heating in the core. We want to stress here that this point is not crucial for our modelling. Comparison of predictions of geo-neutrino production with experimental results at Kamioka is in itself a way of constraining the Earth's energetics, revealing whether the Earth's flow is mainly non radiogenic or significant K has to be hidden in the Earth's interior.

## 4. The Bulk Silicate Earth constraint

So far we have been considering the effect of different geometrical distributions of trace elements in the mantle, for fixed amounts of these elements within it. Actually the BSE model can be exploited so as to obtain tight constraints on the *total* flux produced together from the crust and the mantle. In fact, with BSE fixing the total amount of trace elements inside Earth, geometrical arguments and observational constraints on the crust composition can be used in order to find extreme values of the produced fluxes. As an extension of the previous section, the maximal (minimal) flux is obtained by placing the sources as close (far) as possible to Earth's surface, where the detector is located.

As mentioned in the introduction, the range of BSE Uranium concentrations reported in the literature is between 18 and 23 ppb, corresponding to a total Uranium mass between $m(min)=0.72$ and $m(max)=0.92$ in units of $10^{17}$ kg. In the same units, we estimate that Uranium mass in the crust is between $m_c(min)=0.30$ and $m_c(max)=0.41$, by taking the lowest (highest) concentration reported in the literature *for each layer*, see Table 2 of [3]. The main source of uncertainty is from the abundance in the lower crust, estimated at 0.20 ppm in Ref. [32] and at 1.1 ppm in Ref. [33]. Estimates for the abundance in the upper crust are more concordant, ranging from 2.2 ppm [4] to 2.8 ppm [34]. We remark that, within this approach, the resulting *average* crustal Uranium abundance $a_{cc}$ is in the range 1.3-1.8 ppm, which encompasses all estimates reported in the literature [32, 33, 35, 36] but for that of Ref. [4], $a_{cc}$=0.9 ppm (Table 2).

The highest flux is obtained by assuming the maximal mass in the crust and the maximal allowed mass in the mantle, $m$(max)-$m_c$(max)=0.51, with a uniform distribution inside the mantle, corresponding to $a$=12.8 ppb. On the other hand, the lowest flux corresponds to the minimal mass in the crust and the minimal mass in the mantle, $m$(min)-$m_c$(min)=0.42, with a distribution in the mantle similar to that of PRIM, *i.e.* a strongly depleted mantle with $a_u$=2 ppb down to about 1300 km and a primordial composition beneath.

The predicted fluxes are shown in Table 4 for a few locations of particular interest: the Kamioka mine (33° N 85° E) where KamLAND is operational, the Gran Sasso laboratory (42° N 14° E) where BOREXINO [22] is in preparation, the top of Himalaya (36.N 137. E), which receives the maximal contribution from the crust, and Hawaii (20° N 156° E), a location where the mantle contribution is dominant. At any site the difference between the maximal and the minimal flux predictions are of about 30%, the extreme values being within ±15% from the reference model prediction.

All this shows the power of the BSE constraint. If the total amount of Uranium inside Earth is fixed at $m_{BSE} = (0.8 \pm 0.1) \cdot 10^{17} kg$, then the produced geo-neutrino flux at, *e.g.* Kamioka is:

$$\Phi = (3.7 \pm 0.6) \cdot 10^6 cm^{-2} s^{-1} \quad \text{(full range)} \tag{4}$$

after taking into account *the full range of global* observational uncertainties on Uranium abundances in the crust and uncertainties concerning circulation in the mantle. We insist that the error quoted in Eq.(2) corresponds to a full range of the predicted values. If, following a commonly used rule of thumb, we consider the full range of predictions in (4.1) as a $\pm 3\sigma$ (99.5%) confidence level, we deduce a conventional 1$\sigma$ estimate:

$$\Phi = (3.7 \pm 0.2) \cdot 10^6 cm^{-2} s^{-1} \quad (1\sigma). \tag{5}$$

5. **The effects of local structures**

The main result of the previous section is that - neglecting regional fluctuations - global mass balance provides a precise determination of the geo-neutrino fluxes. We shall compare this precision with uncertainties resulting from fluctuations of the regional geochemical composition.

Indeed the Uranium concentration in the region where the detector is located may be different from the world average and local fluctuations of this highly mobile element are to be envisaged. These variations, although negligible for mass balance, can affect the flux significantly. In other words, geometrical arguments fix the contribution of distant sources and a more detailed geological and geochemical investigation of the region around the detector is needed, the error quoted in Eq. (5) providing a benchmark for the accuracy of the local evaluation. In this respect, let us consider a few examples of practical interest.

*5.1. The contribution from the crust near KamLAND*

It has been estimated that about one half of the geo-neutrino signal is generated within a distance of 500 km from Kamioka, essentially in the Japanese continental shelf. In REF the world averaged upper crust Uranium concentration, $a_{uc} = 2.5\,ppm$, was adopted for Japan. In a recent study of the chemical composition of Japan upper crust [37] more than hundred samples, corresponding to 37 geological groups, have been analyzed. The composition is weighted with the frequency in the geological map and the resulting average abundance is $a_{Jap} = 2.32\,ppm$, which implies a 7.2% reduction of the flux from Japanese upper crust with respect to that estimated in REF. Larger variations occur when rocks are divided according to age or type, see Table 5, and even larger differences are found within each group. All this calls for a detailed geochemical and geophysical study, with the goal of reducing the effect of regional fluctuations to the level of the uncertainty from global geochemical constraints.

*5.2. The subducting slab below the Japan Arc*

As well known, below the Japan islands arc there is a subducting slab originating from the Philippine and Pacific plates. Let us compare the amount of Uranium carried by this plate with that contained in the continental crust of the Japan arc.

Roughly, the Japan crust can be described as a rectangle with area $A=L_1 \cdot L_2 \approx 1800 \cdot 250$ km$^2$ = $4.5 \cdot 10^5$ km$^2$ (Fig. 5). Conrad depth is on the average at $h_1$=18km and Moho discontinuity at $h_2$=36 km [38]. We assume uniform density $\rho$=2.7 ton/m$^3$. Concerning Uranium abundance we take for the upper crust $a_{uc}$=2.3 ppm from [37]. For the lower crust we take $a_{lc}$= 0.6 ppm, an average between largely different estimates. The resulting uranium masses, $m_i = A\, h\, \rho\, a_i$, are reported in Table 6.

The Philippine plate is moving towards the Eurasia plate at about 40 mm/yr and is subducting beneath the southern part of Japan. The Pacific Plate is moving in roughly the same direction at about 80 mm/yr and is subducting beneath the northern half of Japan. The slab is penetrating below Japan with an angle $\alpha \approx 6°$ with respect to the horizontal. This process has been occurring on a time scale $T \approx 10^8$ y. Along this time the slab front has advanced by $D = vT \approx 6000$ km for v=60 mm/yr, the average of the two plates, see Fig. 6. We assume that the slab brings with it oceanic crust, with density $\rho_{oc}$=3 Ton/m$^3$ for a depth $h_3 \approx 10$ km, the Uranium abundance being typical of an oceanic crust, $a_{oc}$=0.1 ppm.

If we assume that the slab keeps its trace elements while subducting, we have just to estimate the amount of Uranium which is contained in the subducting crust below Japan. Its area A' below is slightly larger than that of Japan arc, A' = A cosα ≈ A. For the assumed values of density and depth the mass of the slab is $M_{slab}$ = 1.35 10$^{19}$ kg. The Uranium mass in the subducting crust is thus $m_{slab}$= 1.3 10$^{12}$ kg, a negligible amount as it is about 1/40 of that in the continental crust of Japan.

On the other hand, it is possible that the slab loses Uranium while subducting. As an extreme case, we assume that all Uranium from the subducting crust is dissolved in fluids during dehydration reactions and accumulates in the lower part of the continental crust of Japan, enriching it. Since Japan has been exposed to a slab of length $D \approx 6000$ km, the maximal accumulated Uranium mass is $m_{acc}$=3.2 10$^{13}$kg. This corresponds to an increase of the Uranium abundance in the Japanese lower continental crust, which becomes $a_{lc}$=2 ppm instead of the previously assumed 0.6 ppm. The prediction of the produced flux at Kamioka changes from 3.7 to 4.0 10$^6$ cm$^{-2}$ s$^{-1}$. We remark that this 8% effect has been derived assuming the extreme hypothesis of a complete release.

*5.3. Plumes*

So far we have been considering the mantle as a spherically symmetrical system, whereas, as well known, there are significant inhomogeneities. As an extreme case, let us consider the effect of a plume emerging from the mantle on the vertical of the detector. Clearly what matters is the contrast between the plume and the average mantle, *i.e.* the result essentially depends on the difference between the Uranium abundances in the plume and that in the mantle. For simplicity we assume the detector to be on the top of a cylindrical plume with radius $r_p$, extending down to a depth $h_p$ with uniform density $\rho$ and Uranium abundance $a_p$. The contribution to the geo-neutrino flux from the plume at the detector position $r$ is given by

$$\Phi_p(\vec{r}) = \frac{n_U}{4\pi\mu_U\tau_U}\rho a_p \int_{V_p} d^3r' \frac{1}{|\vec{r}-\vec{r}'|^2} \tag{6}$$

where $V_p$ is the volume of the plume. For the cylindrical plume, the result is:

$$\Phi_p = \frac{A_p}{4}\left\{h_p \ln\left[\frac{h_p^2 + r_p^2}{h_p^2}\right] + 2r_p \, atan\left(\frac{h_p}{r_p}\right)\right\} \tag{7}$$

where $A_p = \frac{n_U \rho a_p}{\mu_U \tau_U}$ is the U-neutrino activity of the plume (i.e. the number of anti-neutrinos produced per unit volume and time from Uranium chain). As shown in Fig. 7, this expression is increasing with the depth of the plume and for a long plume ($h_p >> r_p$) it reduces to the asymptotic value:

$$\Phi^{(as)}_p = \frac{\pi}{4} A_p r_p \tag{8}$$

For a mantle with uniform activity $A_m$, we find from eq. (10) of [15]:

$$\Phi_m \approx \frac{1}{2} A_m R_\oplus \tag{9}$$

where $R_\oplus$ is the Earth's radius. By comparing eqs. (8) and (9) we find that a single long plume just below the detector provides a contribution as large as the whole mantle if its radius $r_p$ and activity $A_p$ satisfy:

$$A_p r_p \approx A_m R_\oplus \tag{10}$$

Since activity is essentially proportional to the element abundance, a similar equation holds for the Uranium abundance in the plume ($a_p$) and the average Uranium abundance in the mantle ($a_m \approx 11.25$ ppb):

$$a_p r_p \approx a_m R_\oplus \tag{11}$$

For $r_p \approx 350$ km, this means $a_p \approx 20\, a_m$, in other words if the Uranium abundance in a plume is 20 times larger than the average Uranium abundance in the mantle, then the plume contribution is comparable to that of the whole mantle. This corresponds to a value exceeding 200 ppb, that is clearly unrealistic. On the other hand, estimates of U abundance in the mantle source of plume-derived OIB magmas with either HIMU or EM isotopic signatures (see [39] and references therein) may be roughly in the order of 30 up to 50 ppb assuming a bulk partition coefficient of 0.002-0.004 for a garnet peridotite assemblage and a nominal melt fraction of 0.01. The U-neutrino flux from a plume with such U abundance is about 20-25% of that from the whole mantle, and thus it might be detectable.

In summary, geo-neutrinos are not useful for measuring the depth of plume columns, however, this could provide an independent way of assessing the existence of plumes, and possibly a measurement of their uranium abundances.

### 4. The required detector size

Let us remark that the signal is originated from neutrinos which maintain the electron flavour in their trip from source to detector, the effective flux being $\Phi_{eff} = \Phi P_{ee}$, where $\Phi$ is the produced flux and $P_{ee}$ is the (distance averaged) survival probability. From the analysis of all available solar and reactor neutrino experiments, one gets [40] $P_{ee} = 0.59 \pm 0.02$. If Uranium geo-neutrinos are detected by means of inverse β-reaction on free hydrogen nuclei $\left(\overline{\nu}_e + p \rightarrow e^+ + n\right)$ the event number is [17]:

$$N = 13.2\varepsilon \left(\frac{\Phi_{eff}}{10^6 cm^{-2} s^{-1}}\right)\left(\frac{N_p}{10^{32}} \frac{t}{yr}\right) \quad (11)$$

where $\varepsilon$ is the detection efficiency, $N_p$ is the number of free protons in the target and $t$ is the measurement time. For a produced flux $\Phi = 4 \cdot 10^6 cm^{-2} s^{-1}$ and $\varepsilon$=80%, one expects 25 events for an exposure of $10^{32}$ protons·yr. Statistical fluctuations will be of order $\sqrt{N}$ if background can be neglected.

In order to reach a 5% accuracy - comparable to that of the global geochemical estimate - one needs an exposure of $16 \cdot 10^{32}$ protons yr, which corresponds to a five-kton detector operating over four years[2]. As a comparison, the data released from KamLAND in 2002 from just six months of data taking correspond to $0.14 \cdot 10^{32}$ protons yr. Several KamLAND size detectors in a few years would be sufficient for collecting the required statistics.

## 5. Concluding remarks

We summarize here the main points of this paper:

1) Uncertainties on the geometrical distribution of trace elements in the mantle (for a fixed mass within it) can change the prediction of the reference model [3] for the geo-neutrino flux from mantle by at most $\pm 8\%$ (full range), the extreme values corresponding to a fully-mixed and to a two-layer model, with primordial abundance below about 1300 km.

2) By using global mass balance for the Bulk Silicate Earth, the predicted flux contribution originating from distant sources in the crust and in the mantle is fixed within $\pm 5\%$ $(1\sigma)$ with respect to the reference model.

3) A detailed geological and geochemical investigation of the region within few hundreds km from the detector has to be performed, for reducing the flux uncertainty from fluctuations of the local abundances to the level of the global geochemical error.

4) A five-kton detector operating over four years at a site relatively far from nuclear power plants can measure the geo-neutrino signal with 5% accuracy. Such a detector is a few times larger than that already operational at Kamioka.

This will provide a crucial test of the Bulk Silicate Earth and a direct estimate of the radiogenic contribution to terrestrial heat. If experiments at Kamioka furnish results close to the predicted minimum values for U and Th, then these elements provide a minor contribution to the earth's energetics; this in turn implies that either Earth's flow is mainly non radiogenic or significant K has to be hidden in the Earth's interior. Alternatively, if experimental results approach the predicted maximum values for U and Th, the Earth's heat flow will be confirmed to derive from the radiogenic contribution.

**Acknowledgments**

---

[2] One kton of liquid scintillator contains approximately $0.8 \cdot 10^{32}$ free protons.

We express our gratitude for useful discussions Dr. C. Bonadiman, L. Carmignani, M. Coltorti, S. Enomoto, K. Inoue, E. Lisi, T. Mitsui, B. Ricci, N. Sleep, A. Suzuki, and F. Villante. The manuscript benefited from constructive reviews and comments by A.N. Onymous and M.Ozima.

This work was partially supported by MIUR (Ministero dell'Istruzione, dell'Università e della Ricerca) under MIUR-PRIN-2003 project "Theoretical Physics of the Nucleus and the Many-Body Systems" and MIUR-PRIN-2002 project "Astroparticle Physics".

**Table 1.** Main radiogenic sources. We report the Q-values, the half lives ($\tau_{1/2}$), the maximal energies ($E_{max}$), heat and anti-neutrino production rates ($\varepsilon_H$ and $\varepsilon_{\bar{\nu}}$) per unit mass for natural isotopic abundances.

| Decay | Q [MeV] | $\tau_{1/2}$ [$10^9$ yr] | $E_{max}$ [MeV] | $\varepsilon_H$ [W/kg] | $\varepsilon_{\bar{\nu}}$ [kg$^{-1}$s$^{-1}$] |
|---|---|---|---|---|---|
| $^{238}U \rightarrow {}^{206}Pb + 8\ {}^4He + 6e + 6\bar{\nu}$ | 51.7 | 4.47 | 3.26 | 0.95·10$^{-4}$ | 7.41·10$^7$ |
| $^{232}Th \rightarrow {}^{208}Pb + 6\ {}^4He + 4e + 4\bar{\nu}$ | 42.8 | 14.0 | 2.25 | 0.27·10$^{-4}$ | 1.63·10$^7$ |
| $^{40}K \rightarrow {}^{40}Ca + e + \bar{\nu}$ | 1.32 | 1.28 | 1.31 | 0.36·10$^{-8}$ | 2.69·10$^4$ |

**Table 2.** Mantle contribution to the produced Uranium geo-neutrino flux. The same Uranium mass in the mantle $m_m = 0.45 \cdot 10^{17}$ kg and abundance in the upper layer $a_u = 6.5$ ppb are assumed in each model.

| Model | Critical depth $h$ [km] | Flux $\Phi$ [$10^6$ cm$^{-2}$s$^{-1}$] |
|---|---|---|
| MIX | 25 | 1.00 |
| REF | 670 | 0.95 |
| PRIM | 1630 | 0.92 |

**Table 3.** Average Uranium abundance in the continental crust.

| Reference | $\langle a_{cc} \rangle$ [ppm] |
|---|---|
| Taylor & Mclennan 1985 | 0.91 |
| Weaver & Tarney 1984 | 1.3 |
| Rudnick & Fountain 1995 | 1.42 |
| Wedephol 1995 | 1.7 |
| Shaw et al. 1986 | 1.8 |
| This work, minimal | 1.3 |
| This work, reference | 1.54 |
| This work, maximal | 1.8 |

**Table 4.** Produced Uranium geo-neutrino fluxes within BSE. Minimal and maximal fluxes are shown, together with the Reference values of [3]. Uranium mass *m* and heat production rate *H* within each layer are also presented.
Units for mass, heat flow and flux are $10^{17}$ kg, TW and $10^6$ cm$^{-2}$ s$^{-1}$ respectively.

|  | *m* | *H* | Himalaya | Gran Sasso | Kamioka | Hawaii |
|---|---|---|---|---|---|---|
|  |  |  |  | Φ |  |  |
| Crust MIN | 0.30 | 2.85 | 4.92 | 2.84 | 2.35 | 0.33 |
| Crust REF | 0.35 | 3.35 | 5.71 | 3.27 | 2.73 | 0.37 |
| Crust MAX | 0.41 | 3.86 | 6.55 | 3.74 | 3.13 | 0.42 |
| Mantle MIN | 0.42 | 3.99 |  | 0.80 |  |  |
| Mantle REF | 0.45 | 4.29 |  | 0.95 |  |  |
| Mantle MAX | 0.51 | 4.84 |  | 1.14 |  |  |
| Total MIN | 0.72 | 6.84 | 5.72 | 3.64 | 3.15 | 1.13 |
| Total REF | 0.80 | 7.64 | 6.66 | 4.22 | 3.68 | 1.32 |
| Total MAX | 0.92 | 8.70 | 7.69 | 4.88 | 4.27 | 1.54 |

**Table 5.** Uranium abundances in the upper continental crust of Japan. Groups correspond to rock's age or type and quoted abundances for each group are area weighted values, from Ref. [27].

| Group | Area % | $a_{uc}$ [ppm] |
|---|---|---|
| Pre-Neogene | 41.7 | 2.20 |
| Pre-Cretaceous | 10.5 | 2.11 |
| Neog-Quat. Igneous rocks | 24.1 | 2.12 |
| Paleog-Cret. Igneous rocks | 14.1 | 3.10 |
| Sedimentary | 39.9 | 2.49 |
| Metamorphic | 21.3 | 1.72 |
| Igneous | 38.4 | 2.48 |
| **Global area weighted average** | **99.6** | **2.32** |

**Table 6.** Estimate for the uranium mass in the continental crust of Japan Islands arc.

| | Crustal Mass [$10^{19}$ Kg] | Uranium abundance [$10^{-6}$] | Uranium Mass [$10^{13}$ kg] |
|---|---|---|---|
| Upper crust | 2.2 | 2.3 | 5.0 |
| Lower crust | 2.2 | 0.6 | 1.3 |
| **Total** | **4.4** | | **6.3** |

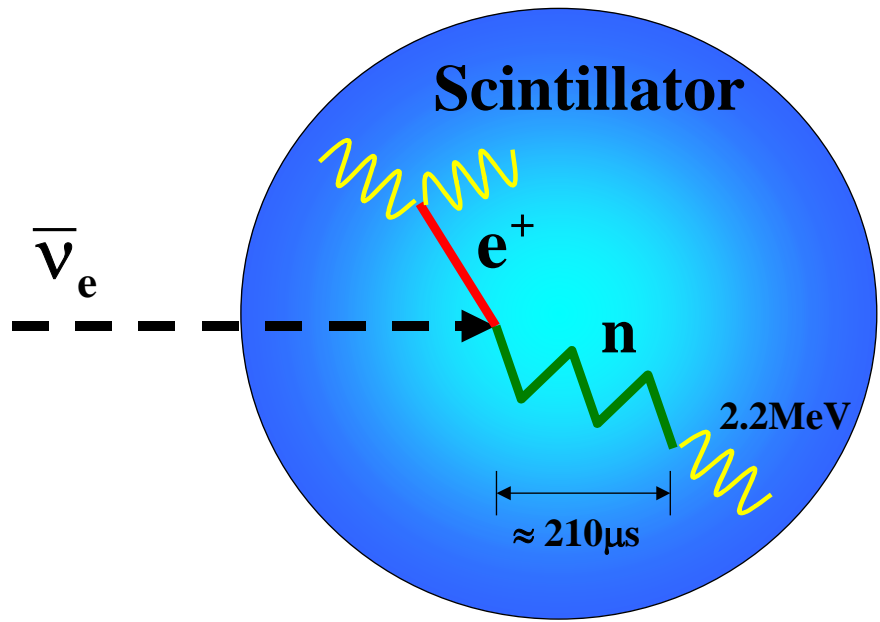

**Fig. 1.** The signature of inverse $\beta$-decay, $\overline{\nu}_e + p \rightarrow e^+ + n$. Energy released in the slowing down of the positron and the two $\gamma$'s from positron annihilation is the prompt signal, followed by the 2.2 MeV $\gamma$-ray from neutron capture on a proton.

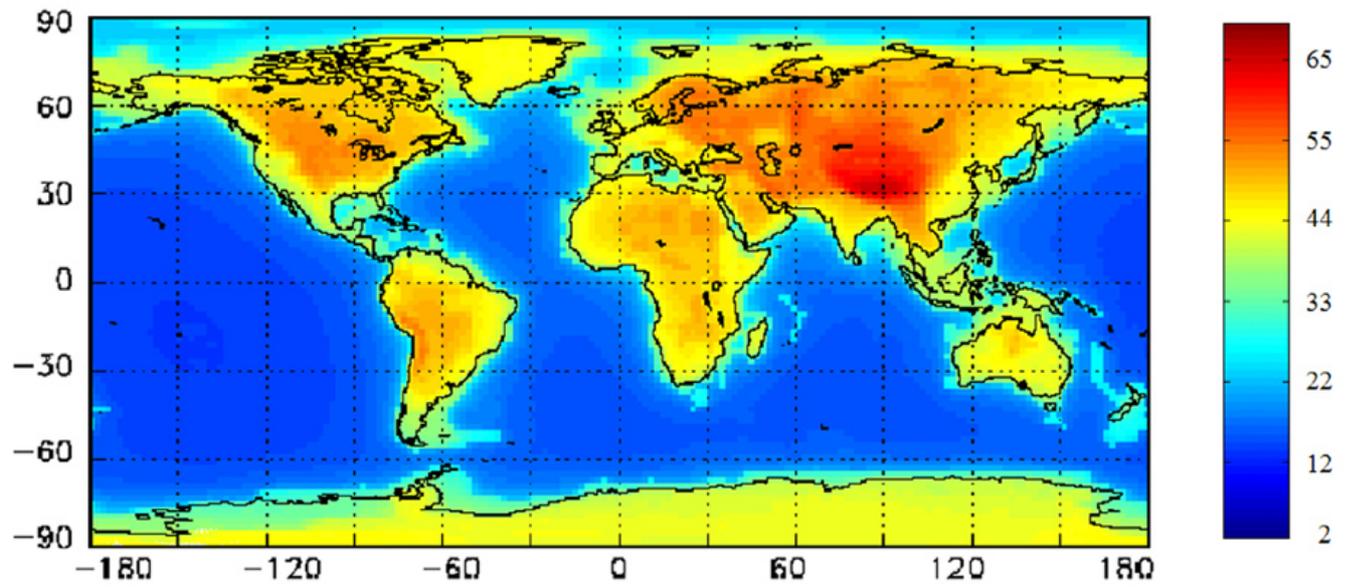

**Fig. 2.** Predicted geo-neutrino events from Uranium and Thorium decay chains, normalized to $10^{32}$ protons yr and 100% efficiency.

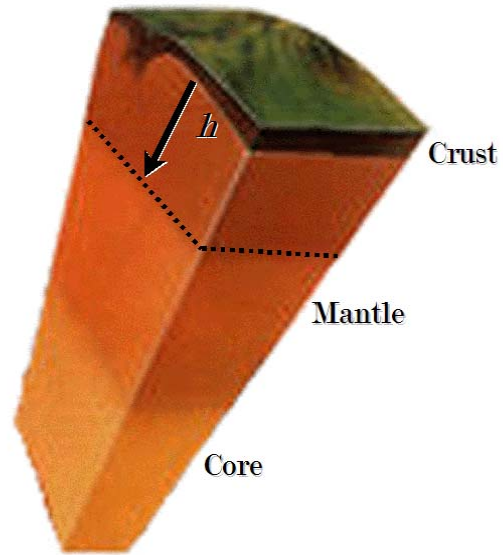

**Fig. 3.** Generic two-reservoir mantle model. Uranium abundance in the upper part is fixed at $a_u = 6.5$ ppb, the critical depth $h$ is a free parameter and the abundance in the lower part $a_l$ is determined for a fixed total Uranium mass in the mantle $m_m = 0.45 \times 10^{17}$ kg.

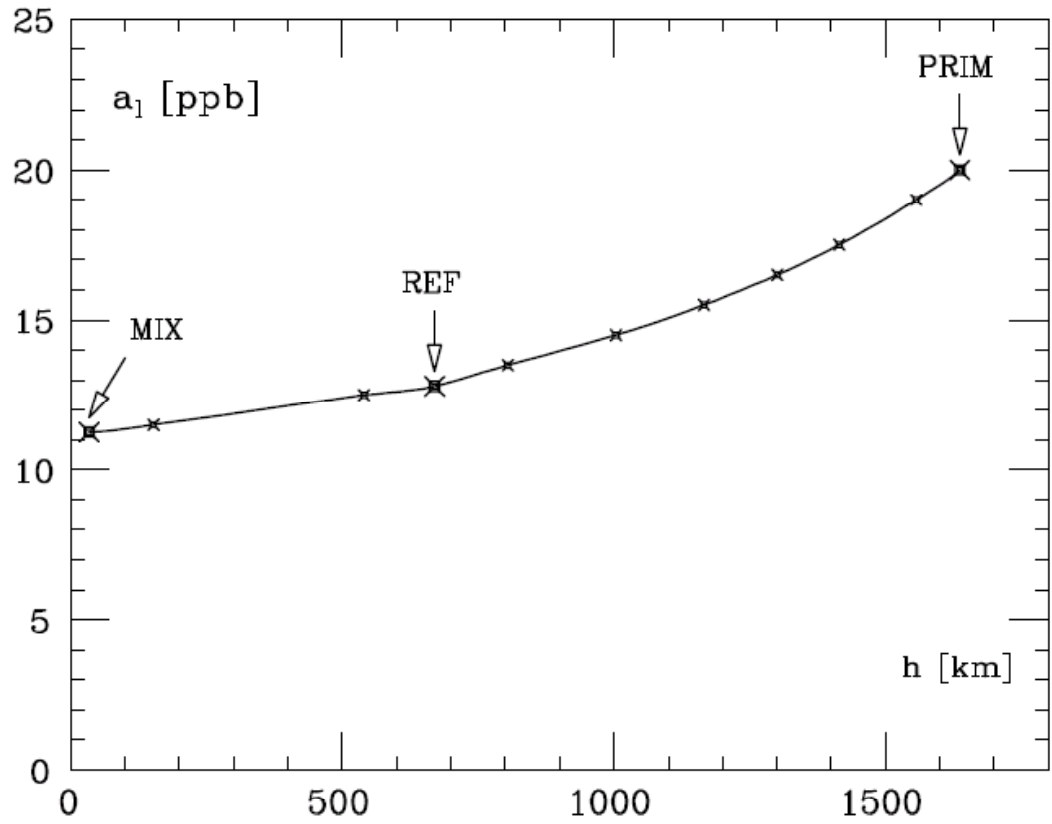

**Fig. 4.** Uranium abundance $a_l$ in the lower part of the mantle as a function of the critical depth $h$ from Earth's surface.

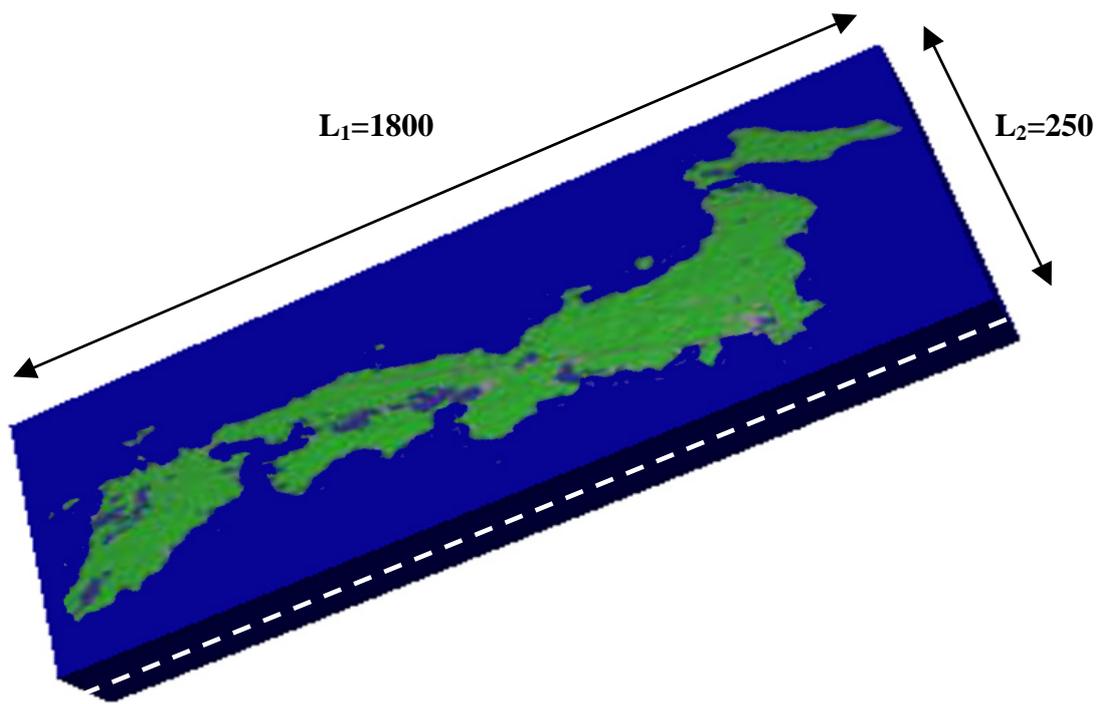

**Fig. 5.** A sketch of the Japan Island Arc.

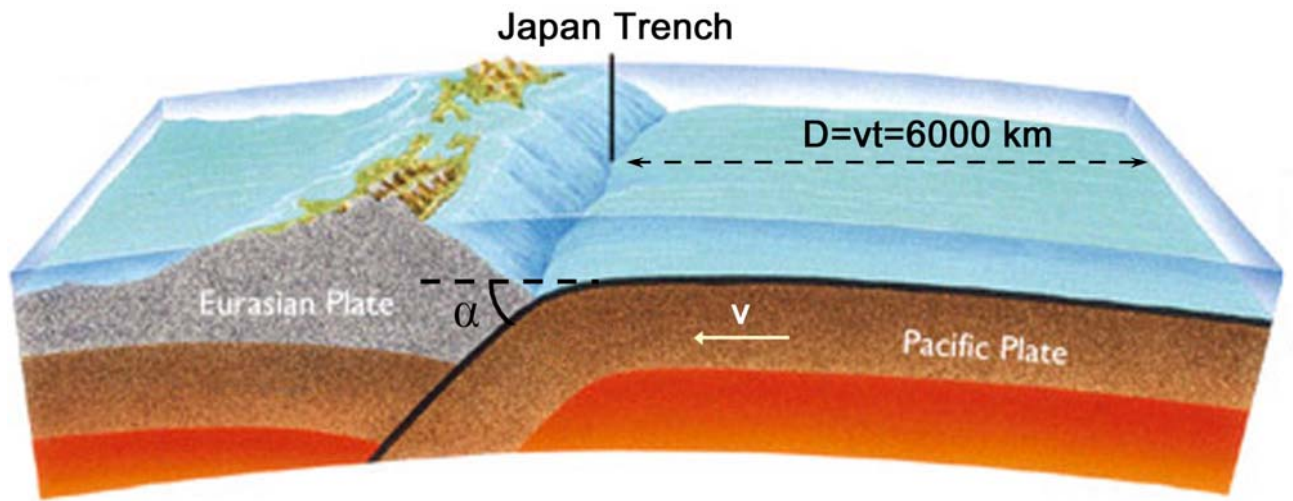

**Fig.6.** A sketch of the Japan arc continental crust and of the subducting slab beneath. The subduction angle is $\alpha \approx 6^0$

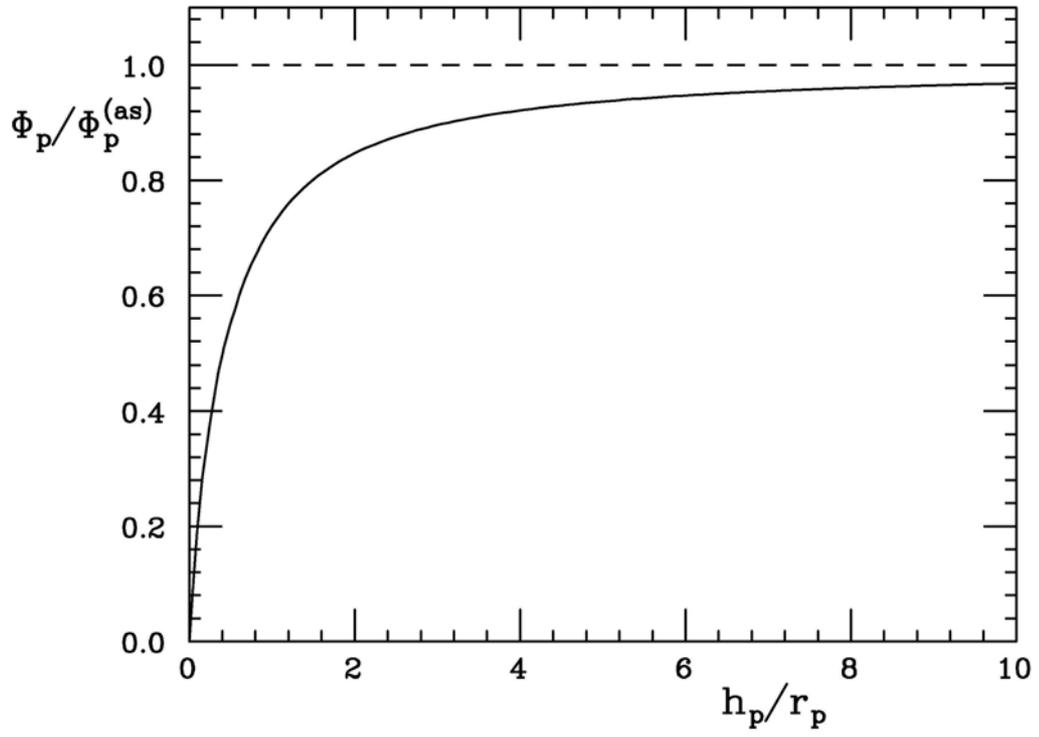

**Fig. 7** The ratio of the plume flux (Eq. 7) to the asymptotic expression (Eq. 8) as a function of $h_p/r_p$